# Design aspects of dual gate GaAs nanowire FET for room temperature charge qubit operation: A study on diameter and gate engineering

Nilayan Paul[1], Basudev Nag Chowdhury[1], Sanatan Chattopadhyay[1,2*]

[1]Department of Electronic Science, University of Calcutta, Kolkata, India.

[2]Center for Research in Nanoscience and Nanotechnology (CRNN), University of Calcutta, Kolkata, India

*E-mail address: scelc@caluniv.ac.in

**Abstract:**

The current work explores a geometrically engineered dual-gate GaAs nanowire FET with state-of-the-art miniaturized dimensions (of nanowire diameter and gate seperation) for high performance charge qubit operation at room temperature. Relevant gate voltages in such device can create two voltage-tunable quantum dots (VTQDs) underneath the gates, as well as can manipulate their eigenstate detuning and the inter-dot coupling to generate superposition, whereas a small drain bias may cause its collapse leading to qubit read-out. Such qubit operations, *i.e.*, 'Initialization', 'Manipulation', and 'Measurement', are theoretically modeled in the present work by developing a second quantization filed operator based Schrodinger-Poisson self-consistent framework coupled to non-equilibrium Green's function (NEGF) formalism. The study shows that the Bloch sphere coverage can be discretized along polar and azimuthal directions by reducing the nanowire diameter and increasing the inter-dot separation respectively, that can be utilized for selective information encoding. The theoretically obtained stability diagrams suggest that downscaled nanowire diameter and increased gate separation sharpen the 'bonding' and 'anti-bonding' states with reduced anticrossing leading to a gradual transformation of the 'hyperbolic' current mapping into a pair of 'straight lines'. However, the dephasing time in the proposed GaAs VTQD-based qubit may be significantly improved (~10 ns to ~100 ns) by scaling down both the nanowire diameter and gate separation to ~5-3 nm. Therefore, the present study suggests an optimization window for geometrical engineering of a dual-gate nanowire FET qubit to achieve a selective coverage of Bloch sphere for particular information encoding, stability diagram of desired resolution with minimum anticrossing, and an extensively improved dephasing time. Most importantly, such device is compatible with the mainstream CMOS technology and can be utilized for large scale implementation by little modification of the state-of-the-art fabrication processes.

**Introduction:**

The practical implementation of quantum computers and quantum information processing requires the development of qubits, a quantum analogue of the classical bits, where quantum superposition is utilized to obtain coherent oscillation between two qubit-states ($|0\rangle$ and $|1\rangle$) within their coherence time [1-2]. Past two decades have exhibited several physical embodiments of such two-level systems, including superconductors with Josephson junctions [3-5], photonic qubits [6-7], trapped-ions [8], semiconductor quantum dot (QD) based charge and spin qubits [9-12], qubits based on defect states in semiconductors [13-14], and topological states of matter [15-16], with each having their own merits and demerits. For instance, photonic qubits operate at room temperature while suffering from scalability issues [17]; trapped-ion qubits are expected to provide extremely long coherence times (~50-600 s), although require costly lasing systems [8]; whereas the qubits based on defect states may offer room temperature operationality [13], however, are challenged by reproducibility constraints [18].

Superconducting qubits, and semiconductor QD based spin and charge qubits on the other hand, have overcome most of the challenges for practical implementation purpose. Such devices have exhibited multifaceted progresses in material manufacturing techniques including the growth of isotopically pure silicon [19], improved operation schemes such as quantum error correction [20] and gate pulsing techniques [21-22]. Such developments have already allowed several industries [23] to commercially implement such qubits [24-27]; however, such qubits are operational at ultra-low temperatures (~mK-µK) that pose the challenges of large-scale production. Nevertheless, the semiconductor QDs [11-12], and especially the double quantum dot (DQD) device based charge qubits [9-10, 28-29], have drawn significant attention due to the desired control over manipulation of their quantum states, as well as their possible compatibility with standard CMOS technology. Such DQD devices operate on the principle of localization of a single excess electron partially in both the QDs, which is manipulated by the inter-dot tunnel coupling **[30-31]** through a series of electrostatically interacting gates. To further improve the manipulation of inter-dot coupling, such DQD systems are coupled to resonant cavities **[32]** to allow photon assisted tunneling or coupled to superconductors to allow Cooper-pair assisted tunneling **[33]**. However, such qubits offer very short dephasing times (~1-10 ns) **[34]**, along with their limitation to operate at extremely low temperatures (~mK), similar to superconducting qubits **[1, 35].**

Therefore, in order to address the present challenges, exploration of alternate avenues by, as little as possible modification of the state-of-the-art device architectures in mainstream CMOS technology is essential **[36-38]**. In this context, a device scheme based on CMOS compatible nanowire field-effect transistor (NWFET), with two separated gates, has recently been proposed for charge qubit operation at room temperature **[39]**. Interestingly, relevant voltages at the two gates of such NWFET can create two VTQDs beneath the gates, tune their eigenstates and control the inter-dot coupling, whereas the collapse of such superposing states is obtained by applying a small bias at the drain **[39]**. Such a device scheme may provide extensive control over quantum states by offering much-needed technological flexibility in terms of device fabrication in practice through optimizing materials and device geometry as well as gate engineering. Further, relevant engineering of geometrical parameters of such device may offer larger or selective coverage of the Bloch sphere, highly resolved stability diagram with desired anticrossing and improved dephasing time. From such perspective, the current work investigates the impact of nanowire (NW) diameter and gate separation of a dual-gate GaAs nanowire FET on its charge qubit operation at room temperature to explore a design window for its superior performance. For such purpose, the entire qubit operation, *i.e.*, 'Initialization', 'Manipulation', and 'Measurement', are theoretically obtained from the simultaneous solution of Schrodinger and Poisson equations in second quantization field operator framework employing Non-Equilibrium Green's Function (NEGF) approach. The effects of varying nanowire diameter and gate separation on qubit manipulation or range over the Bloch sphere are investigated in detail. Subsequently, the relevant stability diagrams are analyzed to study the detuning and anticrossing of QDs during qubit operation. Finally, the engineering aspects of nanowire diameter and gate separation are explored to achieve enhanced-coherence time for superior qubit performance.

**Scheme of device:**

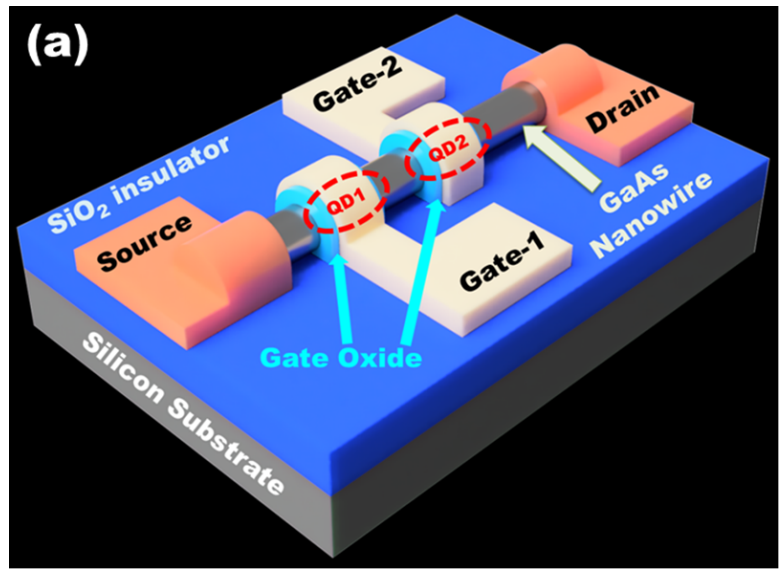

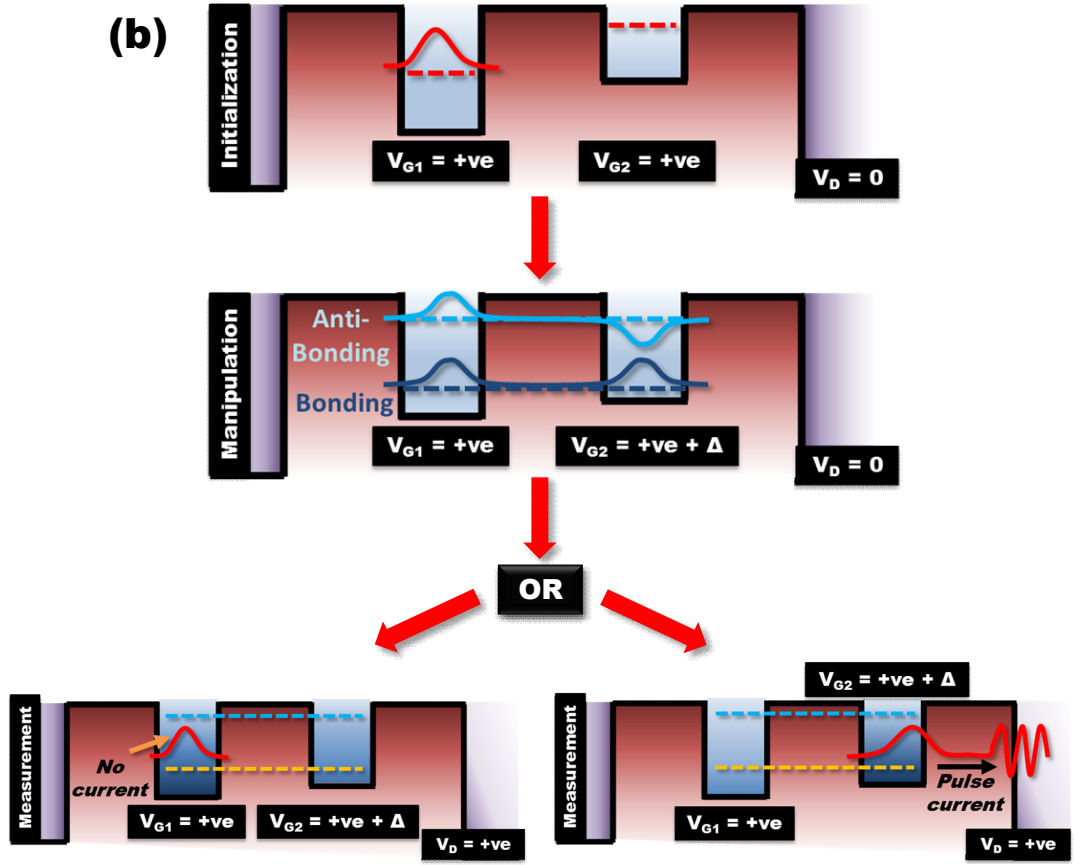

**Fig. 1. (a)** Schematic of dual-gate GaAs nanowire FET considered in the current work. **(b)** Scheme of charge qubit operation in the considered device, depicting 'Initialization', 'Manipulation', and 'Measurement' operations.

A schematic of the proposed dual-gate GaAs nanowire FET device considered in the current work is depicted in Fig. 1(a) where an Ω-gate configuration is assumed. In such a device, the application of appropriate voltages at the two gates leads to the formation of voltage-tunable quantum dots (VTQDs) in the nanowire channel beneath such gates, and the electron occupation of such VTQDs are also controlled by small incremental variation of the applied gate voltages [39]. The nanowire material for the device is considered to be zincblende GaAs (grown along <100>-direction) having a symmetric effective mass tensor of low value (m*~0.06) [40] that can lead to strong quantization of electronic states at room temperature [39, 41]. SiO$_2$ of thickness 2 nm is assumed to be the gate insulator since it has been reported to allow minimum gate leakage by tunneling in nanoscale MOS structures owing to its high electron effective mass (m*~0.4) [42]. The source/drain regions are considered to be degenerate GaAs with a doping concentration of $5\times10^{17}$ cm$^{-3}$ [43]. The total channel length (source-to-drain) of the nanowire FET is considered to be 20 nm to ensure ballistic transport of electrons for minimizing the scattering induced decoherence effects [44]. In order to achieve strong quantum confinement and for the realization of a single-level device at room temperature, the lengths of individual gates are considered to be 3 nm each and the diameters are considered to be less than the excitonic Bohr radius of GaAs (~12 nm) [45], with the minimum and maximum diameters being 5 nm and 10 nm, respectively. It has already been reported that phonon scattering has negligible impact in decohering such highly quantized devices, and the dephasing in present qubit device originates from the fundamental energy-time uncertainty, depending on the inter-dot and reservoir couplings [39]. Finally, to electrically isolate the active device from substrate, the entire device is assumed to be fabricated on an Insulator on Silicon (IOS) platform [46], as shown in Fig. 1(a). At this point it is imperative to mention that all such dimensions are considered according to the feasibility of fabrication as per the state-of-the-art technological accessibility and achievements [47-48]. Fig. 1(b) represents the basic scheme of charge qubit operation in the considered dual-gate nanowire FET device including 'Initialization', 'Manipulation' and 'Measurement'. When appropriate voltages are applied to the gates, *i.e.*, $V_{G1}$ at gate-1 and $V_{G2}$ at gate-2, two 3D-quantized potential wells with single state each ($|L\rangle$ and $|R\rangle$, respectively) are created within the nanowire channel underneath the gates. Depending on the positional asymmetry of such gates and eigenstate detuning, $|L\rangle$-state is initially occupied while $|R\rangle$ remains unoccupied, which results in the 'Initialization' of the charge qubit. Subsequent application of an incremental voltage pulse ($\Delta V_{G2}$ ~mV) on gate-2 manipulates the inter-dot coupling through state manipulation. Through resonant tunneling between the two VTQDs, a superposed state $|\psi\rangle = \alpha(V_{G1}, V_{G2})|L\rangle + \beta(V_{G1}, V_{G2})|R\rangle$ is formed, $|\alpha|^2$ and $|\beta|^2$ being the probabilities of electron localization within the corresponding QDs. In such a condition, the single electron is partially in QD-1 and partially in QD-2 at the same time and thereby forms 'bonding' and 'anti-bonding' states (Fig. 1(b)). The corresponding probabilities to be localized partially to either of the QDs can be manipulated by varying the applied voltage pulse on gate-2 that further modulates the broadening of inter-dot resonance. Such mode of operation is geometrically represented as the rotation of a unit vector on Bloch sphere in polar (θ) direction. However, for a particular value of $V_{G2}$, such unit vector on Bloch sphere will rotate along the azimuthal (ϕ) direction by an incremental variation of $V_{G1}$ [39]. The rotation of unit vector over Bloch sphere in all θ and ϕ directions constitute the 'Manipulation' operation of qubit. At this point, if a small drain bias ($V_D$) is applied, the superposed state will suffer collapse to either $|L\rangle$-state or $|R\rangle$-state. Physically, such a collapse leads to localization of the single electron solely to one QD, resulting in either 'no drain current' (electron in VTQD-1) or a 'pulse current' (electron in VTQD-2 tunneling to the drain), leading to the 'Measurement' of qubit state. It is worthy to mention that in order to aid the 'Initialization' of present qubit device, the source-to-gate-1 distance is fixed at 3 nm, with gate-2 kept distant from the source, such that the electron tunneling probability from source to VTQD-2 becomes negligible. This allows the gate voltages to be defined in such a manner that VTQD-1 can be tuned to hold a single electron from the source while keeping VTQD-2 unoccupied. In order to maintain such conditions, the gate separation is varied in the current work in the range of 3-8 nm. The variation of QD separation eventually affects inter-dot coupling thereby modulating the anti-crossing between 'bonding' and 'anti-bonding' states of the qubit. On the other hand, the variation of nanowire diameter affects the transverse confinement of electrons in the nanowire and thus modulates the energy eigenvalues of the VTQDs. It also modulates the potential profile along the nanowire axis and affects the height of tunnel barriers.

**Theoretical modeling**

The charge qubit operation in the considered dual-gate nanowire FET including the impact of geometrical engineering of such device in terms of nanowire diameter and gate separation are analytically modeled on the basis of NEGF formalism **[39, 41, 49-54]**. The relevant Hamiltonian of the nanowire coupled with source/drain reservoirs is given by,

$$H = \sum_i H_{ISO} c_i^+ c_i + \sum_{i,r=p,q} \left( \zeta_{i,r}^{S/D} c_i^+ C_r + \zeta_{r,i}^{R} * c_i C_r^+ \right) + \sum_p H_S C_p^+ C_p + \sum_q H_D C_q^+ C_q \quad (1)$$

where, $H_{ISO}$ is the Hamiltonian of isolated nanowire channel (*i.e.*, active device region) and $H_{S/D}$ are the Hamiltonians of isolated source/drain regions (*i.e.*, reservoirs); $c_i (c_i^+)$ represents the second quantization annihilation (creation) operator for the electrons (which follow Fermi-Dirac anti-commutation relation) in $i^{th}$ state of the nanowire channel while $C_{p/q} (C_{p/q}^+)$ are that for the electrons in $p^{th}(q^{th})$ state of source/drain. Further, $\zeta_{i,r}^{S/D}$ represents the coupling between the $i^{th}$ state of the nanowire channel with $r^{th}$ state of source/drain. Thus, using the Heisenberg equation of motion, Eq. 1 gives rise to,

$$i\hbar \frac{dc_j}{dt} = H_{ISO} c_j + \sum_p \zeta_{jp}^{S} C_p + \sum_q \zeta_{jq}^{D} C_q \quad (2)$$

for the nanowire, and

$$i\hbar \frac{dC_p}{dt} = H_S C_p + \sum_j \zeta_{pj}^S * c_j \tag{3}$$

$$i\hbar \frac{dC_q}{dt} = H_D C_q + \sum_j \zeta_{qj}^D * c_j \tag{4}$$

for the source and drain, respectively. Using source/drain Green's functions $\left(g^{S/D}\right)$, the equation of motion for electrons in the nanowire channel is obtained to be [53],

$$i\hbar \frac{dc_i(t)}{dt} = H_{ISO} c_i(t) + \sum_{r=p,q} \zeta_{i,r}^{S/D} \left[C_r(t)\right]_{ISO} + \sum_j \int dt' \sum_{r=p,q;m} \zeta_{i,r}^{S/D} g_{r,m}^{S/D}(t,t') \zeta_{m,j}^{S/D} * c_j(t') \tag{5}$$

Subsequently, utilizing the two-time correlated Green's function for the nanowire channel and applying Fourier transform into energy domain, Eq. 5 leads to [50, 53-54],

$$G(E) = \left[E - H_{ISO}(E) - \Sigma_S(E) - \Sigma_D(E)\right]^{-1} \tag{6}$$

where, $\Sigma_S$ and $\Sigma_D$ are the self-energy matrices due to source and drain, respectively, and $G(E)$ is the Green's function for nanowire in the energy domain. At this point, it is imperative to mention that qubit 'Initialization' is incorporated in the model through $\Sigma_S$, whereas $\Sigma_D$ describes the 'Measurement' operation. Interestingly, the preparation of coherent state in the qubit from non-existence of such state, *i.e.*, 'Initialization', and collapsing of such coherent state, *i.e.*, 'Measurement', are both non-unitary evolution of the quantum state, and therefore, consistently described in the present formalism by such non-Hermitian self-energy matrices. However, the impact of gate voltages, which do not decohere the unitary evolution, are incorporated in the model through $H_{ISO}$, using 'coupled mode space' approach [55-56]. In fact, the electrons in nanowire FET are confined in the transverse direction (denoted by x, y-coordinates), while they are free to move along the nanowire axis (denoted by z-coordinate) until localized gate voltages are applied. Therefore, $H_{ISO}(E)$ can be written as [50],

$$H_{ISO}(E) = \sum_n E_n - \frac{\hbar^2}{2}\left(\frac{\partial}{\partial z}\left(\frac{1}{m_z^*}\frac{\partial}{\partial z}\right)\right) - e\varphi(z) \qquad (7)$$

where, $E_n$ is the n$^{th}$ mode due to transverse confinement, and obtained from [39, 50],

$$H_T|n_T\rangle = \left[-\frac{\hbar^2}{2}\sum_{j,k}\partial_j\left(\frac{1}{m_{j,k}^*}\partial_k\right) + (-e)\varphi_T(x,y;z)\right]|n_T\rangle = E_n|n_T\rangle \qquad (8)$$

In Eq. 7, $\varphi(z)$ denotes the conduction band potential along the nanowire axis, while $\varphi_T(x,y;z)$ in Eq. 8 is the conduction band potential (for any particular z-coordinate) in the direction transverse to the nanowire axis; and $m_{j,k}^*$ corresponds to the (j, k)$^{th}$ element of electron effective mass tensor corresponding to such transverse directions. It is imperative to mention that, in the present qubit, the values of $E_n$ and their gaps are dependent on the nanowire diameter. Thus, such diameters can be selected to increase the energy gap between ground state and 1$^{st}$ excited state to make the impact of phonon scattering negligible, even at room temperature [39]. Further, in such case, the 'single state' qubit operation requires only one transverse mode, that significantly improves the computational efficiency in 'coupled mode space' approach. Once the Green's function for active device is calculated, the electron correlation function for occupied states in energy domain can be obtained as [39, 53],

$$[n(E)] = [G(E)][\Sigma_S^{In}(E) + \Sigma_D^{In}(E)][G^+(E)] \tag{9}$$

where, $\Sigma_{S/D}^{In}$ represents the carrier inflow from source/drain to nanowire, given by,

$$[\Sigma_{S/D}^{In}(E)] = [\zeta][i(\Sigma_{S/D}(E) - \Sigma_{S/D}^+(E))f_{S/D}(E)][\zeta^+] = [\zeta][\Gamma_{S/D}(E)f_{S/D}(E)][\zeta^+] \tag{10}$$

with, $\Gamma_{S/D}$ and $f_{S/D}$ describing the source/drain induced broadening and Fermi-Dirac distribution functions, respectively. Therefore, the occupied local density of states (LDOS) for the present nanowire FET device consisting of two gate-controlled VTQDs may be written as,

$$D(E) = \frac{2}{2\pi a}n(E) = \frac{i}{\pi a}\begin{bmatrix}(G(E)[\Sigma_S(E) - \Sigma_S^+(E)]G^+(E))f_S(E) \\ +(G(E)[\Sigma_D(E) - \Sigma_D^+(E)]G^+(E))f_D(E)\end{bmatrix} \tag{11}$$

where, '$a$' is the grid spacing along the nanowire axis. It is worth mentioning that Eq. 11 considers Kramer's degeneracy to consider the probability of two electrons with different spin occupying the same energy state **[39]**. However, whether a single state will be occupied by two electrons or a single electron, is determined by the alignment of VTQD energy eigenstate with respect to the source/drain Fermi levels. Further, it must be noted that although spin-degeneracy is considered in the present work, spin-orbit coupling is not considered since its effect in GaAs quantum structures is reported to be observed only at very low temperatures **[57-59]**.

The carrier profile $(n(z))$ along nanowire axis is obtained by integrating such occupied LDOS over all possible energy values, while its normalization over positional coordinates (*i.e.*, along nanowire channel) yields the probability density $(|\psi(z)|^2)$ of electron within the device. Such carrier profile, since the Poisson's equation for a nanowire FET with Ω-gate configuration given by [46],

$$\left[\frac{d^2}{dz^2} - \frac{2}{R^2}\frac{\varepsilon_{Ox}}{\varepsilon_{NW}}\frac{1}{\ln\left(1+\frac{t_{Ox}}{R}\right)}\right](\varphi(z) - V_G(z)) = \frac{e}{\varepsilon_{NW}(\pi-\theta)R^2}n(z) \qquad (12)$$

leads to the resultant potential distribution along the nanowire channel $(\varphi(z))$ including the two gate-defined VTQDs. It may be noted that in Eq. 12, $\varepsilon_{NW}$ and $\varepsilon_{Ox}$ are the permittivity of nanowire material (GaAs) and gate oxide (SiO$_2$), respectively; $t_{Ox}$ is the thickness of gate oxide and $R$ is nanowire radius; $V_G(z) = V_{G1(2)}(z)$ throughout gate-1(2), while zero elsewhere; and $\theta = \cos^{-1}(R/(R+t_{Ox}))$ for the Ω-gate configuration without any insertion of nanowire into the IOS [46]. It is interesting to note that Eq. 12 shows the dependence of potential variation along the nanowire channel on its diameter. Therefore, not only does the choice of nanowire diameter affect the energy eigenstates of the VTQDs (Eq. 7 and 8), but it also influences the potential profile of the nanowire, thereby manipulating the heights of inter-dot tunnel barrier as well as source/drain barriers. At this point, it is worth mentioning that the output potential of Eq. 12 and the input potential of isolated Hamiltonian in Eq. 6 needs to be self-consistent from computational perspective for the quantum-electrostatic conditions to be fulfilled simultaneously, from the physical point-of-view. This is obtained in the current work by iterative computation, and once self-consistency is achieved within the desired limit of accuracy, the pulse current at drain is calculated from Landauer formula [60],

$$I = \frac{2e}{h}\int dE\, T(E)[f_S(E) - f_D(E-V_D)] \qquad (13)$$

where, $T(E) = Trace[\Gamma_S(E)G(E)\Gamma_D(E)G^+(E)]$ is the transmission coefficient.

It is of significant importance to mention at this point that the VTQDs with very few numbers of atoms have no defined temperature (since no statistical average is valid). Thus, the ambient temperature can affect the VTQDs only through coupling with the reservoirs (*i.e.*, source/drain). For instance, at the 'Initialization' and 'Manipulation' operation modes, the effect of temperature intrudes predominantly through the coupling with source, while during 'Measurement', majorly the drain induces the impact of temperature. Thus, the coupling strength of VTQDs with source/drain determines the role and influence of temperature in the present qubit device. This is incorporated in the present model through source/drain Fermi-Dirac distributions, associated with self-consistent occupied LDOS (Eq. 11) and pulse current (Eq. 13). Eq. 13 provides the amplitude of pulse current, its oscillatory decay due to coherent oscillation of the qubit along with the natural dephasing is obtained to be **[39, 53]**,

$$I = I_0 \left( F.T._{E \to t} [G_{ISO}(E) \Sigma_D(E) G(E)] \right) \tag{14}$$

where, $(F.T._{E \to t})$ indicates Fourier transform from energy domain to time domain and $G_{ISO}(E)$ represents the Green's function for isolated nanowire channel when the coupling with source/drain is negligible. It is interesting to note from a modeling point of view that the Green's function $G(E)$ in Eq. 14 contains both real (Hermitian) and imaginary (non-Hermitian) components (see Eq. 6). The Hermitian component of Green's function is responsible for the coherent evolution of superposed state, which is physically achieved when $\Sigma_{S/D}(E) \to 0$, *i.e.*, $G(E) \to G_{ISO}(E)$. However, the application of drain bias $(V_D)$, results in non-Hermitian contribution corresponding to 'Measurement' operation of the qubit (as discussed on Eq. 6). The unitary evolution combined with the non-Hermitian 'Measurement' is responsible for the oscillatory decay of the current at the drain, which is also discussed in detail from physical point of view in [39]. The schematic of such self-consistent procedure for modeling the charge qubit operation in present dual-gate nanowire FET is shown in Fig. 2.

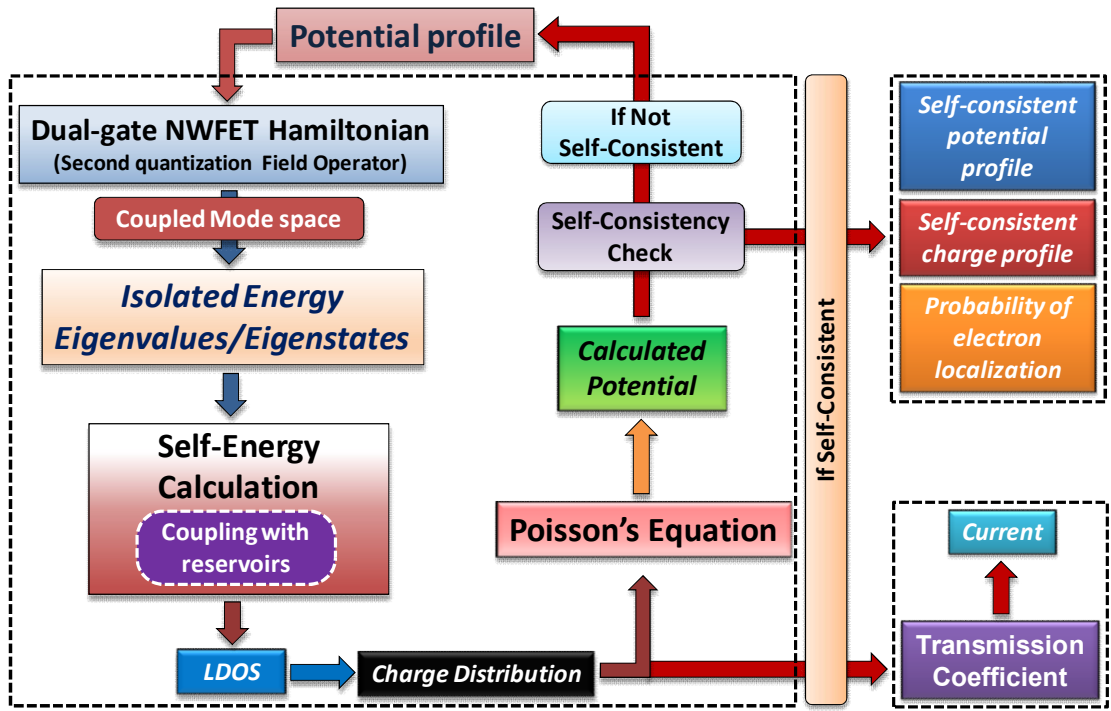

**Fig. 2** Schematic of the self-consistent procedure for modeling the dual-gate nanowire FET, and corresponding charge qubit operation at room temperature.

**Results and Discussion:**

To develop a comprehensive understanding on the impact of nanowire diameter and gate separation on the charge qubit operation of dual-gate nanowire FET, the manipulation of superposed states is analyzed through Bloch sphere coverage of such states, as shown in Fig. 3(a)-(d). Such figures comparatively represent the polar and azimuthal angle variations of the superposition over Bloch sphere for nanowire diameters of 5 nm and 10 nm with gate separations of 3 nm and 8 nm, respectively.

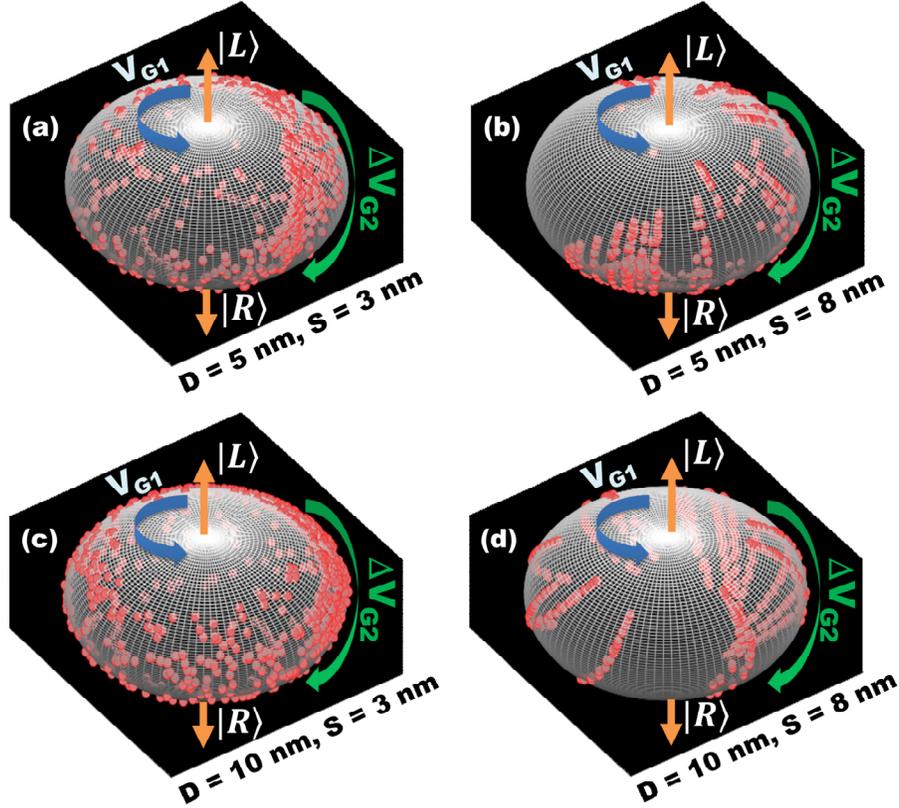

**Fig. 3** Bloch spheres representing the superposed states $|\psi\rangle = \cos\left(\frac{\theta}{2}\right)|L\rangle + e^{i\phi}\sin\left(\frac{\theta}{2}\right)|R\rangle$ by varying $V_{G1}$ and $V_{G2}$ at $V_D = 0$ for **(a)** nanowire diameter (D) = 5 nm and separation (S) = 3 nm; **(b)** nanowire diameter (D) = 5 nm and separation (S) = 8 nm; **(c)** nanowire diameter (D) = 10 nm and separation (S) = 3 nm; **(d)** nanowire diameter (D) = 10 nm and separation (S) = 8 nm. The dots on the surface of the spheres map the position of Bloch vector for combinations of applied gate voltages, where inter-dot resonance tunneling occurs. The trajectory of such Bloch vector corresponds to the manipulation of the superposed state during qubit operation.

The qubit superposed state may be expressed in spherical polar coordinates (ignoring the global phase) as,

$$|\psi\rangle = \cos\left(\frac{\theta}{2}\right)|L\rangle + e^{i\phi}\sin\left(\frac{\theta}{2}\right)|R\rangle \qquad (15)$$

where, $\theta$ is a measure of the probability of an electron to be localized in the $|L\rangle$ and $|R\rangle$ states, which are represented by the north and south poles, respectively, while $\phi$ is the relative phase between such states. For the considered device, rotation of qubit state from north to south pole (θ-rotation) on Bloch sphere is performed by varying pulsed $V_{G2}$ for a fixed $V_{G1}$, whereas the relative phase between $|L\rangle$ and $|R\rangle$ is manipulated ($\phi$-rotation) by varying $V_{G1}$, keeping $V_{G2}$ fixed [39]. It is apparent from Fig. 3(a)-(b), as well as Fig. 3(c)-(d) that the increase in inter-dot separation from 3 nm to 8 nm (*i.e.*, weakening of coupling) results in discrete coverage of the Bloch sphere in $\phi$-direction. This is attributed to the fact that the phase difference of two VTQDs (*i.e.*, $\phi$) depends on electron momentum, which gets discretized for weak inter-dot coupling resulting in almost 0-D quantization [61]. Such reduction in Bloch sphere coverage eventually allows much lesser but selective information to be encoded in the qubit device during its operation. On the other hand, downscaling of nanowire diameter leads to the omission of few information in the θ-direction of Bloch sphere due to stronger confinement leading to energetically discretized probability (since θ measures probability) of resonance tunneling from left QD to right QD.

The impact of geometrical engineering of the considered device in terms of nanowire diameter and gate separation on room temperature charge qubit operation are further analyzed by investigating the charge-stability diagrams (Fig. 4 and 5). Fig. 4(a)-(f) show the plots of qubit current ($I_0$) as the 'Measurement' results for 'Manipulation' of two gate voltages ($V_{G1}$ & $V_{G2}$) in the devices of different nanowire diameters (5 nm – 10 nm) with identical inter-dot separation (3 nm). Such charge stability diagrams are manifestation of the splitting of the single electronic state of two QDs into two energy states: 'bonding' and 'anti-bonding', due to inter-dot resonant tunnel coupling. Such energy states are functions of the detuning ($\Delta\varepsilon = \varepsilon_1 \sim \varepsilon_2$) between isolated states ($\varepsilon_1$ and $\varepsilon_2$) of the two QDs and their anticrossing ($\tau$), and conventionally described by

$$E^* = \frac{\varepsilon_1 + \varepsilon_2}{2} \pm \frac{1}{2}\sqrt{(\Delta\varepsilon)^2 + 4|\tau|^2} \qquad (16)$$

as a simplified model **[9, 29]**. However, in practice, such states exhibit energy level broadening due to finite coupling between the VTQDs themselves (during coherent manipulation of superposed state), as well as between the VTQDs and source/drain reservoirs (during Initialization/Measurement operations of the qubit). Such level broadening is consistently incorporated in the present formalism in relevance to experimental observations **[28-30]** through self-energy matrices (see Eq. 6).

It is evident from Fig. 4(a)-(f), that the downscaling of nanowire diameter results in a decrease in the anticrossing between 'bonding' and 'anti-bonding' states. For instance, the anticrossing decreases from ~11 meV in the device with 10 nm nanowire diameter device to ~6 meV for a 5 nm diameter device. This is attributed to stronger confinement in smaller dimensions, leading to level sharpening, which further weakens the inter-dot coupling. Moreover, the increase in nanowire diameter also reduces the source/drain barrier heights, leading to additional source/drain-induced-broadening **[54, 60]** of the 'bonding' and 'anti-bonding' energy states, that is observed as a 'smearing' of such states near the anticrossing line (see Fig. 4(c)-(f)). At this point, it is imperative to mention that a 'sharpened' current line distinguishing the charge states ((0,0), (1,0), (0,1), (1,1)) is necessary for superior qubit performance. Therefore, the source/drain induced level broadening accompanying the strong coupling regime of the VTQDs may be detrimental to qubit performance, since the distinguishability of charge qubit states may get lost. Thus, smaller nanowire diameters are possibly the suitable choice for superior qubit performance. Further, a lower operational temperature may reduce the level broadening, and hence may improve the qubit stability behavior.

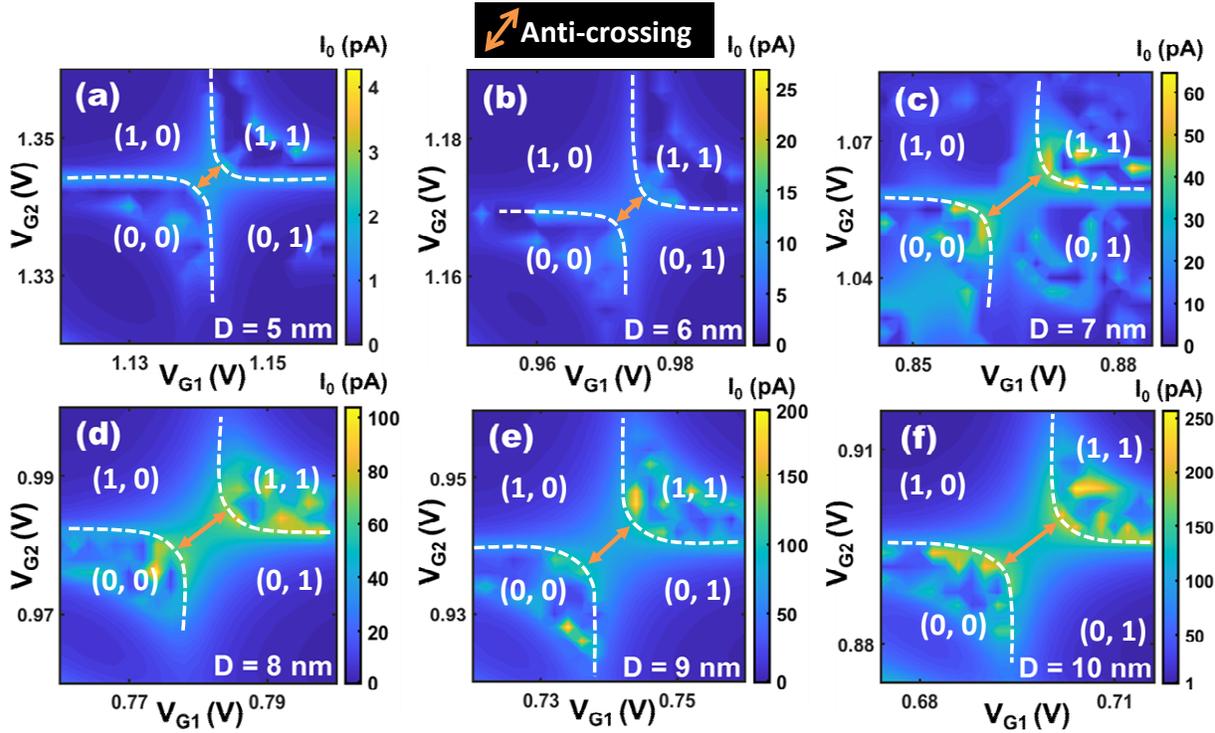

**Fig. 4** Charge stability diagrams for device with different nanowire diameter (D), while keeping gate separation (S) = 3 nm and gate lengths fixed at 3 nm. The diameter increases from 5 nm (shown in (a)) to 10 nm (shown in (f)). The increase in nanowire diameter results in the increase in coupling between the VTQDs, and manifests in the charge stability diagram through the enhancement of hyperbolic nature at anticrossing point, while increasing the magnitude of anticrossing ($\tau$).

On the other hand, the impact of gate-separation, *i.e.*, inter-dot distance, on charge qubit operation of the dual-gate nanowire FET is shown in Fig. 5(a)-(f), where, keeping the nanowire diameter fixed (5 nm), the separation between VTQDs is varied from 3 nm to 8 nm. It is apparent from the figures that inter-dot separation plays a significant role in the qubit manipulation, resulting in a drastic alteration of the overall nature of the stability diagram. Physically, the increase in inter-dot separation leads to reduced strength of inter-dot coupling due to larger width of inter-dot tunnel barrier.

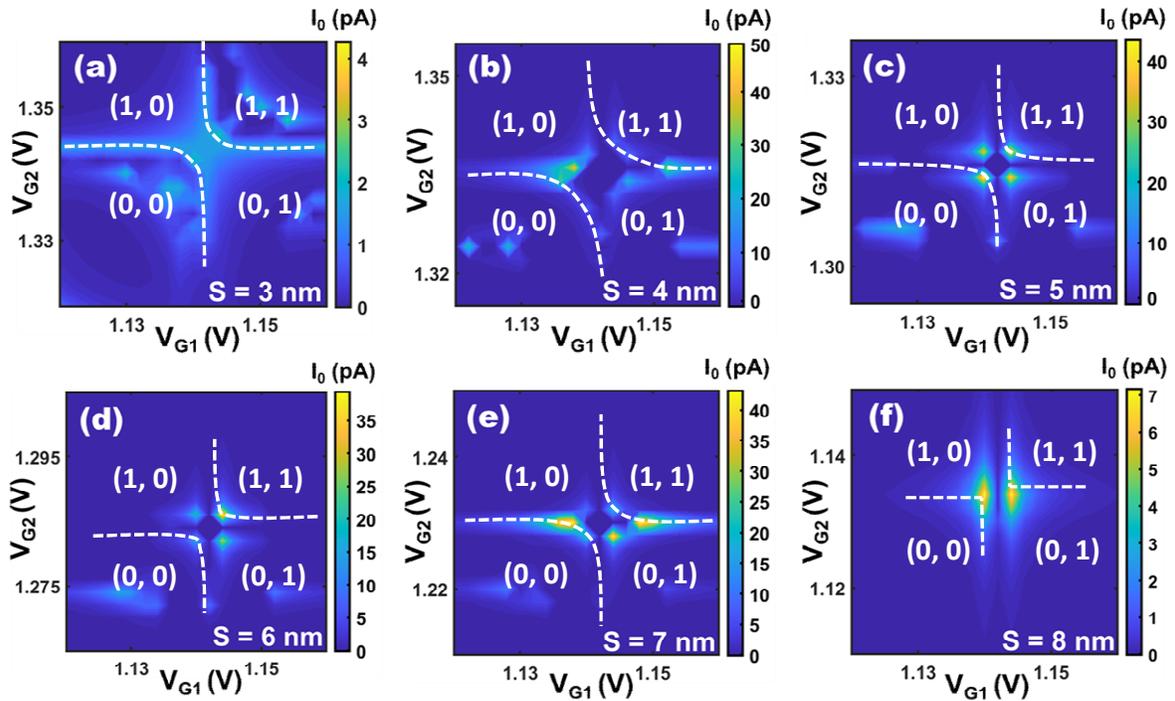

**Fig. 5** Charge stability diagrams for device with changing gate separation (S), keeping nanowire diameter (D) fixed at 5 nm, and gate lengths fixed at 3 nm. In the figures, the gate separation (QD separation) is increased from 3 nm (Fig. (a)) to 8 nm (Fig. (f)). The increase in gate

separation leads to a decrease in inter-dot coupling and reflected in the corresponding stability diagrams.

The weakening of inter-dot coupling results in sharper resonance tunneling between the VTQDs during manipulation; however, much weaker coupling (*i.e.*, due to large separation) may significantly reduce the tunneling probability. Consequently, with increasing inter-dot separation the current line in stability diagram exhibits sharpening along with reduction of anticrossing, and the 'hyperbolic' nature of current gradually transforms into 'straight line', as apparent from the plots of Fig. 5(a)-(f). Such weakening of coupling results in further splitting of the 'bonding' and 'anti-bonding' states in (Fig. 5(b)-(d)), which can be understood conceptually from Eq. 16 of the simplified model mentioned earlier. For an extremely weak inter-dot coupling, the anticrossing energy gets so small that the energies of superposed state may be obtained using binomial approximation as,

$$E^* = \frac{\varepsilon_1 + \varepsilon_2}{2} \pm \frac{\Delta\varepsilon}{2}\left(1 + 2\frac{|\tau|^2}{(\Delta\varepsilon)^2}\right) \tag{17}$$

Thus, at the vicinity of ideal resonant condition (*i.e.*, zero detuning), the energy has four possible solutions:

$$E^* = \begin{cases} \varepsilon_1 \pm \dfrac{|\tau|^2}{|\Delta\varepsilon|} \\ \varepsilon_2 \pm \dfrac{|\tau|^2}{|\Delta\varepsilon|} \end{cases} \tag{18}$$

This, in turn, leads to the development of four current lines/spots associated with 'bonding/anti-bonding' (see Fig. 5(c)-(e)). However, when the inter-dot separation is further increased (>7 nm), such splitting disappears in the direction of $V_{G2}$ (see Fig. 5(f)), which is attributed to the significant coupling of VTQD-2 with drain due to their reduced separation (~3 nm). At this point, it is worthy to mention that the stability diagrams corresponding to such moderate and weak inter-dot coupling in double quantum dot devices, where the 'hyperbolic' nature of current/conductivity mapping transforms to 'straight line' nature, are reported in a number of experimental results **[28-30, 36, 62-66]**.

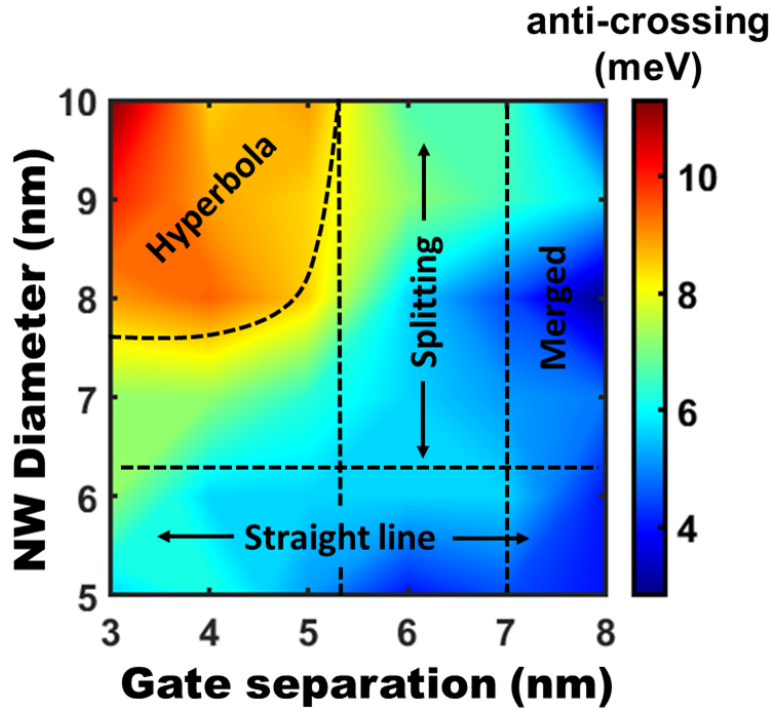

**Fig. 6** Contour plot of anticrossing energy for combination of different nanowire diameters and gate separations. The values of anticrossing energy are observed to increase with the increase in nanowire diameter, with a maximum value of ~11 meV for device with 10 nm nanowire diameter and 3 nm gate separation. Anticrossing energy decreases with increasing gate

separation to a minimum of ~4 meV for device with 5 nm nanowire diameter and 8 nm gate separation.

The resultant impacts of nanowire diameter and gate separation on qubit anticrossing energy are summarized in a relevant contour plot as shown in Fig. 6. It is apparent from Fig. 6 that the impact of inter-dot separation on anticrossing energy is dominant for larger nanowire diameters; for instance, the anticrossing decreases from ~11 meV to ~6 meV for an increase in inter-dot separation from 3 nm to 8 nm in the device with 10 nm nanowire diameter, while from ~6 meV to only ~4 meV in a 5 nm diameter device. This may be attributed to the stronger confinement in smaller dimensions itself weakening the coupling. At this point it is further worthy to mention that although the thermal energy of room temperature (*i.e.,* ~26 meV) is higher than such anticrossing energy values, it cannot cause overlapping of the 'bonding' and 'anti-bonding' states due to weak coupling of the QDs with reservoirs. However, for 8 nm gate separation, the coupling of VTQD-2 with drain leads to significant overlap (see Fig. 5(f)). It is further imperative to note that anticrossing energy in the charge stability diagrams is a measure of inter-dot coupling strength, where, in fact, the weak coupling regime favors the efficient manipulation of inter-dot bonding in a wide qualitative range: 'ionic' to 'covalent' bond of an artificial molecule [9].

On the other hand, the dephasing time being a measure of sustenance of coherent oscillations between the 'bonding' and 'anti-bonding' states, is one of the key performance parameters of a qubit [1, 35]. Such dephasing time is extracted in the present work from the decay profile of oscillatory pulse current at drain during measurement (by applying a small $V_D$) [39]. The variation of output signal in time for continuous qubit measurements in the devices with different nanowire diameters and different gate separations, are depicted in Fig. 7(a) and (b), respectively. It may be worthy to note that such output signal at the drain is a direct consequence of the manipulation of superposed state in time domain by the application of $\Delta V_{G2}$ as pulses, with its duration less than the overall pulse repetition time for 'Initialization', which is obtained to be ~336.5 ns in the present device [39].

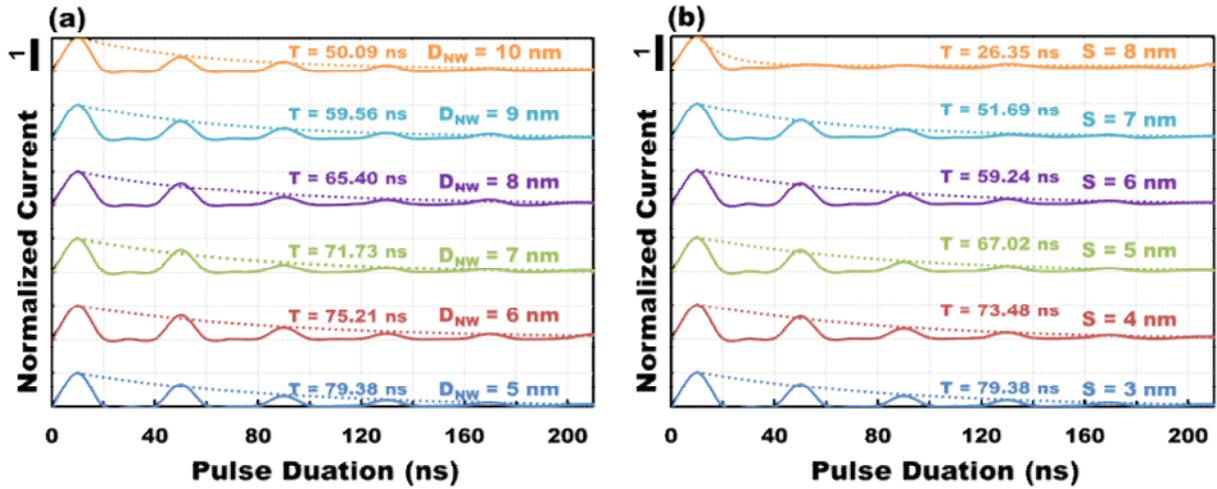

**Fig. 7 (a)** Pulse current at drain for devices with varying nanowire diameter (D), ranging from 5 nm to 10 nm, keeping inter-dot separation (S) fixed at 3 nm; **(b)** Pulse current at drain for devices with varying inter-dot separation (S), from 3 nm to 8 nm, keeping nanowire diameter (D) fixed at 5 nm. The variation of such current is obtained at the operating points of the devices corresponding to the bonding/anti-bonding states at anticrossing point and defined by appropriate $V_{G1} - V_{G2}$ combinations.

It is observed from the plots of Fig. 7(a) and (b), that the increase in both nanowire diameter and gate-separation (inter-dot separation) result in faster decay of the pulse current indicating quicker dephasing of the qubit. For instance, the dephasing time for a device with 10 nm nanowire diameter and 3 nm inter-dot separation is found to be ~50 ns, i.e., much lower than the ~80 ns for a 5 nm diameter device with identical inter-dot separation. In the latter device, the dephasing time decreases with increasing inter-dot separation up to 8 nm. Evidently, a slower dephasing or a longer coherence time can be achieved by improving the inter-dot resonant tunneling. Therefore, smaller diameter leading to stronger confinement and closely spaced QDs enhances the coupling strength which increase the qubit dephasing time. Further, there is an additional decohering effect that appears from drain coupling during 'Measurement', which can be minimized by applying a small bias at drain as well as by keeping QD-2 distant from the drain. It is worth noting that decreased dephasing times degrade qubit performance, and therefore in the current work, a geometrical engineering of the dual-gate nanowire FET based qubit is proposed in terms of nanowire diameter and inter-dot separation to maximize such time scales. This is represented as a contour plot of dephasing times for all combinations of nanowire diameter and inter-dot separation considered in the present device architecture and shown in Fig. 8.

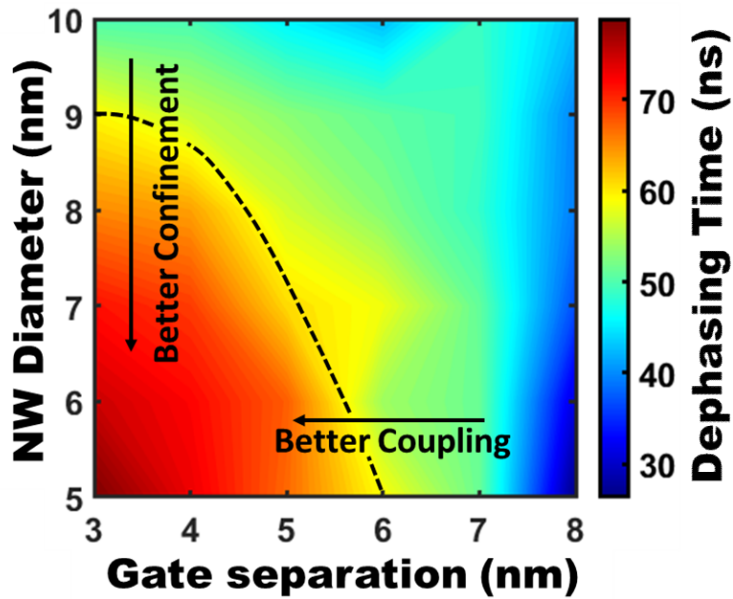

**Fig. 8** Contour plot of decoherence time for charge qubit operation. Both plots have been obtained for varying combinations of nanowire diameter and inter-dot separation, with minimum and maximum nanowire (NW) diameters being 5 nm and 10 nm, respectively; minimum and maximum Gate (QD) separations are 3 nm and 8 nm, respectively.

It is interesting to note that unlike the anticrossing energy (Fig. 6), which suffers a counteracting impact of nanowire diameter and inter-dot separation, the dephasing time is dependent on both nanowire diameter as well as inter-dot separation in the same manner. Thus, to obtain a desired range of anti-crossing and dephasing time, the nanowire diameter and gate separation need to be optimized. In this context, Fig. 8 in conjunction with Fig. 6, provides a window for high performance qubit where long coherence times can be achieved while keeping relatively lower anti-crossing through geometrical engineering of the dual-gate nanowire FET in terms of nanowire diameter and gate separation.

**Conclusions:**

In conclusion, the current work analyzes the impact of geometrical engineering of dual-gate GaAs nanowire FETs in terms of nanowire diameter and gate separation for achieving desired charge qubit operation at room temperature. The fundamental qubit operational modes, *i.e.*, 'Initialization', 'Manipulation', and 'Measurement', in such device are theoretically modeled by developing a Schrodinger-Poisson self-consistent framework based NEGF formalism. It is observed from the results that the Bloch sphere coverage is discretized, in polar direction for downscaling of the nanowire diameter, whereas along azimuthal direction for increasing the inter-dot separation. This may provide an optimization window of device geometry to obtain selective spaces on Bloch sphere for desired information encoding. Further, reduced nanowire diameter leads to sharpen the 'bonding' and 'anti-bonding' states along with decreasing their anticrossing due to stronger quantum confinement. On the other hand, increased gate separation weakens the inter-dot coupling thereby sharpening the resonance tunneling between VTQDs, which is manifested in the stability diagram as a gradual transformation of the 'hyperbolic' nature of current mapping into a pair of 'straight lines'. A further increase of gate separation results in splitting of such pair of current lines into four lines/spots, which again get merged for very high inter-dot separation owing to the drain induced level broadening. Interestingly, the dephasing time in such GaAs VTQD-based qubit may be enhanced from ~10 ns to ~100 ns by downscaling both the nanowire diameter and gate separation to ~5-3 nm. Therefore, the study suggests a design window for relevant engineering of geometrical parameters of the dual-gate NWFET device that may offer a larger or selective coverage of the Bloch sphere, highly resolved stability diagram with desired anticrossing, and significantly improved dephasing time. The detail analyses of the physical origins of such qubit performance parameters also indicate that appropriate material engineering of the present device architecture may provide further scope to enhance such performance for room/moderate temperature charge qubit operation for large scale implementation.

## Acknowledgement

N. Paul would like to acknowledge the University Grants Commission (UGC), India for financial support as a Junior Research Fellow through University of Calcutta. The authors would like to acknowledge WBDITE, DST PURSE, Center of Excellence (COE), and Centre for Research in Nanoscience and Nanotechnology (CRNN) for providing infrastructural support to conduct this work.